\documentclass[a4paper,11pt]{article}
\usepackage{jheppub}
\pdfoutput=1

\usepackage{cancel}
\usepackage[normalem]{ulem}
\usepackage{float}
\usepackage{autobreak}
\usepackage{graphicx}
\usepackage{appendix}
\usepackage{amsfonts}
\usepackage{comment}
\usepackage{multirow}

\usepackage{color,hyperref}
\usepackage[T1]{fontenc}
\usepackage{tocloft}
\usepackage[normalem]{ulem}
\usepackage{subcaption}
\usepackage{array}
\usepackage{enumitem}
\usepackage{extarrows}
\usepackage{mathtools}
\usepackage{slashed} 

\newcommand{\nn}{\nonumber\\}

\newcommand{\df}{{\rm d}}
\def\Dm1{{{\delta(1-z)}}}

\def\g0#1DY{{g_{0#1}^{DY}}}

\def\LogmW1{{{\ln (1-\omega)}}}

\newcommand{\overbar}[1]{\,\overline{\!{#1}}}
\newcommand{\Nbar}{\overbar{N}}
\newcommand{\gbar}{\overbar{g}}
\newcommand{\as}{a_s}
\newcommand{\muf}{\mu_F}
\newcommand{\mur}{\mu_R}

\newcommand{\siglogg}{\widehat{\sigma}^{(0)}_{gg}}
\newcommand{\sigloggEFT}{\widehat{\sigma}^{(0),\text{EFT}}_{gg}}

\newcommand{\MH}{M_H}
\newcommand{\MZ}{M_Z}
\newcommand{\sw}{s_w}
\newcommand{\cw}{c_w}

\newcommand{\zb}{{\bar{z}}}
\newcommand{\Lw}{L_\omega}
\newcommand{\wb}{\overbar{\omega}}
\newcommand{\NLL}{\overline{\text{NLL}}}

\newcommand{\eq}[1]{Eq.\ (\ref{#1})}
\newcommand{\fig}[1]{Fig.\ \ref{#1}}
\newcommand{\tab}[1]{Table\ \ref{#1}}
\newcommand{\sect}[1]{Section\ \ref{#1}}

\note{TTK-24-58, P3H-24-100, IPPP/24/81}

\title{{Soft gluon resummation for gluon fusion $ZH$ production}}
\author[a]{Goutam Das,}
\emailAdd{goutam@physik.rwth-aachen.de}
\affiliation[a]{Institut f{\"u}r Theoretische Teilchenphysik und Kosmologie,\\ RWTH Aachen University, D-52056 Aachen, Germany}
\author[b]{Chinmoy Dey,}
\emailAdd{d.chinmoy@iitg.ac.in}
\affiliation[b]{Department of Physics, Indian Institute of Technology Guwahati,\\ Guwahati-781039, Assam, India}
\author[b]{M. C. Kumar}
\emailAdd{mckumar@iitg.ac.in}
\author[c]{and Kajal Samanta}
\emailAdd{kajal.samanta@durham.ac.uk}
\affiliation[c]{Institute for Particle Physics Phenomenology,\\ Durham University, Durham DH1 3LE, United Kingdom}

\abstract{
We examine the effects of soft gluons on Higgs boson production in association with a $Z$ boson at the Large Hadron Collider (LHC).
Utilizing the universal cusp anomalous dimensions and splitting kernels, we analyze effects of soft gluons on the gluon fusion 
$ZH$ process, focusing on the total production cross section as well as the invariant mass distribution at the next-to-leading 
logarithmic level.
Additionally, we estimate the next-to-soft effects on this subprocess to the same level of accuracy. 
A detailed phenomenological 
analysis is performed for the $13.6$ TeV LHC. Finally, combining these results with those from other subprocesses, we provide 
comprehensive predictions for the $Z H$ production cross section and the invariant mass distribution that will be valuable for 
comparison with experimental data from the upcoming LHC run as well as the future hadron colliders.
}

\begin{document}



\maketitle

\flushbottom

\section{Introduction} \label{sec:introduction}
The discovery of the Higgs boson \cite{Englert:1964et, Higgs:1964pj, Guralnik:1964eu} at the Large Hadron Collider 
(LHC) \cite{ATLAS:2012yve, CMS:2012qbp} marks a significant milestone in particle physics. Following this breakthrough, 
the primary focus of the LHC has been the precise measurement of the Higgs boson's properties and couplings. Such efforts 
not only deepen our understanding of the Higgs boson itself but also open new avenues for exploring physics beyond the 
standard model (BSM). Many BSM scenarios predict weak couplings to the Higgs boson at collider energies in the TeV range. 
Evidence of such interactions may manifest as deviations in total cross sections or distributions for Higgs boson processes. 
Therefore, precise experimental measurements, combined with accurate theoretical predictions, are critical for identifying 
potential BSM signatures. The associated Higgs production with a vector boson at the LHC is of particular importance in 
probing Higgs coupling to the weak gauge bosons. Furthermore, it is crucial to constrain the sign of the top-quark Yukawa 
coupling and examine its $CP$ structure \cite{Englert:2013vua, Hespel:2015zea, Goncalves:2015mfa}.  Keeping in mind the importance 
of this process, the ATLAS and CMS experiments are actively conducting precise measurements for this 
process \cite{ATLAS:2019yhn, ATLAS:2020jwz, ATLAS:2020fcp, CMS:2020gsy, ATLAS:2021wqh, CMS:2024ksn, CMS:2024srp, CMS:2024tdk, CMS:2024fkb, CMS:2023vzh, ATLAS:2024yzu}. Precise theoretical predictions are thus essential to complement the improved experimental 
results for this channel.

At the LHC, the Higgs boson is produced in association with $Z$ boson primarily through the  Drell-Yan (DY) type subprocess 
$(q\bar{q} \rightarrow Z^* \rightarrow ZH)$ at the leading order (LO) in the strong coupling ($\alpha_s$) expansion. The 
higher-order quantum chromodynamics (QCD) corrections for this channel thus closely follow the DY process and have been 
computed up to next-to-next-to-leading order (N2LO) 
\cite{Han:1991ia,Brein:2003wg,Brein:2004ue,Brein:2012ne,Ferrera:2014lca,Campbell:2016jau,Ferrera:2017zex,Harlander:2018yio} 
and recently to next-to-next-to-next-to-leading order (N3LO) \cite{Baglio:2022wzu} accuracy. The perturbative series for the
DY-type contribution converges very quickly with next-to-leading order (NLO) amounting to around $30\%$ correction to LO, 
whereas N2LO and N3LO receive around $3\%$ and $-0.8\%$ corrections, respectively relative to preceding orders. Additionally, the 
conventional theoretical scale uncertainty reduces to subpercent level. The $ZH$ process also gets a contribution from the 
bottom quark annihilation through the $t$-channel diagrams at the LHC. This involves a bottom Yukawa coupling ($y_b$) and its 
contribution is found \cite{Ahmed:2019udm} to be at the subpercent level. Furthermore, the electroWeak (EW) effects for this 
DY-type process have been computed to NLO \cite{Ciccolini:2003jy} and a correction of around $-5\%$ was observed compared to the
NLO QCD result. When fiducial cuts are applied, these corrections can grow significantly,  reaching $-10\%$ to $-20\%$ of the 
NLO QCD distributions \cite{Denner:2011id}.

Compared to the DY process, the $ZH$ production process receives additional QCD contributions starting from N2LO. One such contribution arises from Feynman diagrams in which the Higgs boson is radiated off a massive top-quark loop in the quark–antiquark annihilation subprocess. The impact of this contribution has been estimated in \cite{Brein:2011vx} to be below $3\%$.

Another class of subprocesses that first appears at N2LO, i.e. at ${\cal O}(\alpha_s^2)$, is the gluon fusion channel for 
$ZH$ production. Although the gluon fusion subprocess is suppressed by two powers of the strong coupling constant compared to the quark–antiquark annihilation channel, this suppression is largely compensated by the sizable gluon luminosity at the LHC. Consequently, from N2LO onwards, the gluon fusion contribution becomes phenomenologically significant.
Due to its importance, the gluon-fusion subprocess has been extensively investigated in the literature. At leading order (which contributes at ${\cal O}(\alpha_s^2)$), it is a loop-induced process in which the $ZH$ final state is produced via massive quark loops comprising both triangle and box topologies (see Fig.~2 of \cite{Kniehl:2011aa}) with the top quark providing the dominant contribution. The gluon fusion subprocess at LO has been shown \cite{Dicus:1988yh,Kniehl:1990iva,Kniehl:1990zu,Kniehl:2011aa} to yield a correction of about $7\%$  relative to the NLO QCD DY-type contribution. In contrast, the N2LO QCD DY-type correction amounts to only about $3\%$  relative to NLO DY. The scale uncertainty of the gluon-fusion contribution at leading order is approximately $25\%$.

Several efforts have been made to further improve the accuracy of the gluon-initiated subprocess by calculating its NLO ($\mathcal{O}(\alpha_s^3)$) contribution. This calculation involves two-loop amplitudes with multiple kinematic scales, including three masses: the Higgs boson mass ($M_H$), the $Z$ boson mass ($M_Z$), and the top-quark mass ($m_t$) as well as two Mandelstam variables. The presence of such a large number of scales substantially increases the computational complexity. The problem, however, becomes considerably simpler in the infinite top-mass [effective field theory (EFT)] limit, where the NLO correction has been computed \cite{Altenkamp:2012sx}. Nevertheless, ongoing efforts aim to account for the full top-mass dependence. The NLO real and virtual contributions have been evaluated using an asymptotic expansion in the inverse top-quark mass \cite{Hasselhuhn:2016rqt}. In addition, two-loop amplitudes have been computed through high-energy and large-$m_t$ expansions employing the Padé approximation \cite{Davies:2020drs}, as well as through a transverse-momentum expansion \cite{Alasfar:2021ppe}. The combined effects of the high-energy and transverse-momentum expansions have been further investigated in Refs.~\cite{Bellafronte:2022jmo, Degrassi:2022mro}. The NLO cross section and invariant mass distribution have also been computed by incorporating the full top-quark mass dependence, supplemented by a small-mass expansion in $M_H$ and $M_Z$ \cite{Wang:2021rxu}. More recently, two-loop virtual corrections including full top-mass effects have been evaluated \cite{Chen:2020gae}, followed by a determination of the NLO cross section with complete top-mass dependence \cite{Chen:2022rua}. The resulting NLO total cross section is found to increase by approximately $100\%$ relative to the leading-order prediction, while the associated scale uncertainty is reduced to about $15\%$. In contrast, differential distributions, such as the Higgs boson transverse-momentum spectrum within the fiducial phase space, can receive corrections as large as an order of magnitude.

The fixed-order results for this subprocess still suffer from the large threshold logarithms arising from soft gluons emission. These large logarithms need to be resummed to have reliable predictions. The size of the NLO corrections indicates that this subprocess will receive significant contributions from the threshold logarithms similar to the Higgs case. Resummation of these large soft-virtual (SV)  logarithms is well established in the literature \cite{Sterman:1986aj, Catani:1989ne, Catani:1990rp,Kidonakis:1997gm,Kidonakis:2003tx,Moch:2005ba,Laenen:2005uz,Kidonakis:2005kz,Ravindran:2005vv,Ravindran:2006cg,Idilbi:2006dg,Becher:2006mr,Ahmed:2020nci} 
%
%
and they have been applied to many colorless processes \cite{Catani:2003zt,Moch:2005ky,deFlorian:2007sr,Kidonakis:2007ww,Catani:2014uta,Bonvini:2014joa,
Ahmed:2015sna,Schmidt:2015cea,Ahmed:2016otz,Bonvini:2016frm,Kidonakis:2017dmh,AH:2019phz,Das:2019btv,Das:2019bxi,Das:2020gie,Das:2020pzo,AH:2020cok,AH:2022elh,Banerjee:2024xdh} 
%
%
leading to improved predictions for inclusive cross sections and invariant mass distributions. Recently, efforts were made to incorporate also the next-to-soft (NSV) threshold effects
\cite{Kidonakis:1996aq,Moch:2009hr,Soar:2009yh,Bonocore:2015esa,DelDuca:2017twk,Beneke:2018gvs,Bahjat-Abbas:2019fqa,Beneke:2019oqx,Beneke:2019mua,Moult:2019mog,Liu:2020tzd,vanBeekveld:2019cks,Das:2020adl,AH:2020iki,AH:2021vdc,vanBeekveld:2021hhv,AH:2022lpp,Beneke:2022obx,Liu:2022ajh,Sterman:2022lki,Pal:2023vec}.
%

%
For the $ZH$ production in the DY-type channel, the effects of soft gluons have been estimated in \cite{Kumar:2014uwa,Das:2022zie,Das:2024auk} 
%
%
to next-to-next-to-next-to-leading logarithmic (N3LL) accuracy matched to N3LO fixed order in QCD. A better perturbative convergence has been observed for threshold resummation for invariant mass distribution of $ZH$ pair.
For the gluon fusion $Z H$ process, the impact of soft gluon effects on the total cross section has been investigated in \cite{Harlander:2014wda} to next-to-leading logarithmic (NLL) accuracy matched to NLO in the EFT approximation using two different threshold definitions, namely the $M$ approach and the $Q$ approach. In the $M$ approach, the threshold is defined in terms of the total mass of the $Z H$ system, while in the $Q$ approach, it is defined using the invariant mass of the $Z H$ system. Although both approaches yield results that are consistent within the scale uncertainties, the $Q$ approach exhibits better perturbative stability and slightly reduced scale dependence.

The goal of this paper is to improve the gluon fusion $ZH$ process by incorporating soft and next-to-soft gluon resummation 
for both the total cross section and the invariant mass distribution of the $ZH$ pair. We work in the exact Born-improved gluon 
fusion channel, which has been shown to work effectively for the  Higgs case, and we expect similar behavior for the $ZH$ process.
For NSV resummation, we closely follow the approach outlined in \cite{AH:2020iki,AH:2021vdc}. Additionally, we present the 
complete result at $\mathcal{O}(\alpha_s^3)$, improved with SV threshold resummation from both quark and gluon channels for 
$ZH$ production.

The article is organized as follows: In \sect{sec:theory}, we introduce the key theoretical formulas and present the coefficients 
required for performing SV and NSV resummation up to the necessary order. In \sect{sec:numerics}, we provide a phenomenological 
study for the gluon fusion subprocess, combining it with the DY-type contributions to present complete results for $pp$ 
collisions at $\mathcal{O}(\alpha_s^3)$ accuracy.  Finally, we conclude in \sect{sec:conclusion}.

\section{Theoretical Framework} \label{sec:theory}
The hadronic cross section for $Z H$ production in proton collision is provided in QCD factorization as,
\begin{align}\label{eq:had-xsect}
Q^2 \frac{\df \sigma}{\df Q^2}  
=  
\sum_{a,b}\int_0^1 \df x_1\int_0^1 \df x_2 \,\,f_{a}(x_1,\mu_F^2)\,
f_{b}(x_2,\mu_F^2) 
\int_0^1 \df z~ \delta \left(\tau-zx_1 x_2\right)
Q^2 \frac{\df \widehat\sigma_{ab}(z,\muf^2)}{\df Q^2} \, ,

\end{align}
where $f_{a,b}$ are the parton distribution functions (PDFs) for parton $a,b$ in the incoming hadrons
and $\widehat\sigma_{ab}$ is the partonic coefficient function.  
The hadronic and partonic threshold variables $\tau=Q^2/S$ 
and $ z= Q^2/\widehat{s}$ are defined in terms of respective center-of-mass energies $S$ and $\widehat{s}$. Here $Q$ is 
the invariant mass of the $Z H$ system and $\muf$ is the factorisation scale.
The partonic coefficient function can be decomposed (suppressing all scale dependencies) as,
\begin{align}\label{eq:PARTONIC-DECOMPOSE}
Q^2 \frac{\df \widehat\sigma_{ab}(z)}{\df Q^2}
& =
\widehat{\sigma}^{(0)}_{ab}(Q^2) \Big( 
\delta_{b {a}}\Delta_{ab}^{\rm SV}\left(z\right) 
+ \Delta_{ab}^{\rm REG}\left(z\right)
\Big) \,.

\end{align}
The term $\Delta_{ab}^{\rm SV}$ appears only for the diagonal subprocesses ($gg,q\bar{q}$) and is known as the soft-virtual
(SV) partonic coefficient and it captures the leading singular terms in the $z \equiv 1-\zb \to 1$ limit. 
The $\Delta_{ab}^{\rm REG}$ term, on the other hand, contains subleading or regular contributions in the variable $z$. 
In general, they follow the expansion in the strong coupling at the renormalization scale   
$\alpha_s(\mur^2) \equiv 4 \pi a_s(\mur^2)$ as,
\begin{align}\label{eq:SVREG-Expansion}
\Delta^{\rm SV}_{ab}(z) 
&= \sum_{n=0}^{\infty}\as^n(\mur^2) ~\delta_{ab}
\left( 
\Delta^{(n)}_{\delta} ~\delta(\zb) + 
\sum_{k=0}^{2n-1}  \Delta^{(n)}_{{\cal D}_k} ~{\cal D}_k(\zb) 
\right)\,, ~ 
\text{with } ab \in \{gg, q\bar{q} \} \,,
\\
\Delta^{\rm REG}_{ab}(z) 
&= 
\Delta^{\rm NSV}_{ab}(z) 
+
\sum_{n=0}^{\infty}\as^n(\mur^2)
\Delta_{ab}^{(n)}(z)
= \sum_{n=0}^{\infty}\as^n(\mur^2)
\left( 
\sum_{k=0}^{2n-1} \Delta_{ab,{\ln}_k}^{(n)} ~\ln^k(\zb) 
+
\Delta_{ab}^{(n)}(z)
\right)\,,
\nonumber

\end{align}
where $\delta(\zb)$ is the Dirac delta distribution and ${\cal D}_k(\zb) \equiv \left[\ln^k(\zb)/\zb\right]_+$ are 
the plus distributions. At leading order, all the coefficients vanish except  $\Delta^{(0)}_{\delta} = 1$.
Here $\Delta_{ab,{\ln}_k}^{(n)} $ are the NSV coefficients which get contributions from both diagonal as 
well as off-diagonal channels. The last term in the above expression $\Delta_{ab}^{(n)}(z)$ vanishes in the soft limit 
$\zb\equiv 1-z \to 0$.
Note that the singular SV part of the partonic coefficient has a universal structure which gets contributions from the 
underlying hard form factor,
mass factorization kernels 
\cite{Moch:2004pa,Vogt:2004mw} 
and soft radiations 
\cite{Ravindran:2005vv,
Ravindran:2006cg,
Sudakov:1954sw,
Mueller:1979ih,
Collins:1980ih,
Sen:1981sd}. All of these are 
infrared divergent which, however, when regularized 
and combined give finite contributions.

The Born normalization factor $\widehat{\sigma}^{(0)}_{ab}$ in \eq{eq:PARTONIC-DECOMPOSE} takes the following 
form for the $q\bar{q}$ subprocess:
\begin{align}
	\widehat {\sigma}_{q\bar{q}}^{(0)}(Q^2) 
	&= 
	\bigg(\frac{\pi \alpha^2}{n_c S}\bigg)\Bigg[ 
		\frac{M_Z^2\lambda^{1/2}(Q^2,M_H^2,\MZ^2)
		\bigg( 1 + \frac{\lambda(Q^2,M_H^2,\MZ^2)}{12 M_Z^2/Q^2}\bigg)}{(Q^2-M_Z^2)^2c_w^4s_w^4}
        \left((g_q^V)^2+(g_q^A)^2 \right) \Bigg]\,,
        
\end{align}
%
with $ g_a^A = -\frac{1}{2} T_a^3 , ~g_a^V = \frac{1}{2} T_a^3  - s_w^2 Q_a $, $Q_a$ being the electric charge and 
$T_a^3$ being the weak isospin of the fermions. 
Here, $\alpha$ is the fine structure constant, $s_w$ and $c_w$ are the sine and cosine of the Weinberg angle respectively, 
and $n_c=3$ in QCD.  The function 
$\lambda$ is defined as $\lambda(z,y,x) = 	\left(1-x/z- y/z \right)^2 -4 x y /z^2$.

For the gluon fusion subprocess, in the infinite top-mass limit, the Born factor takes the form \cite{Altenkamp:2012sx},
\begin{align}
\sigloggEFT(Q^2)&= 
  \left(\frac{\sqrt{\pi}~\as(\mur^2)\alpha }{16\sw^2\cw^2}\right)^2
 \,
\left( \frac{Q^2}{\MZ^4} \right) \lambda^{\frac{3}{2}}(Q^2,\MH^2,\MZ^2) \,.

\label{eq:M0square} 
\end{align}
In the infinite top-mass approximation, the full NLO correction to the gluon fusion subprocess is known \cite{Altenkamp:2012sx}. 
In particular, the SV and NSV coefficients up to NLO are given as,
\begin{align}
\Delta^{(1)}_{{\delta}} =& ~
                        \left(\frac{56}{27} + 8 \zeta_2\right) C_A
                        +\frac{64}{9}T_F n_f  
                        + \left(\frac{46}{9} C_A-2\beta_0 \right)\ln\left(\frac{\muf^2}{Q^2}\right)
                        -2 \beta_0        \ln\left(\frac{\muf^2}{\mur^2}\right)
                        \nn
                        &\hspace{2em}+\frac{\widehat\sigma_\text{(virt,red)}}{\as(\mur^2)\sigloggEFT(Q^2)} \,,
\nn
\Delta^{(1)}_{{\cal D}_0} =& -\Delta_{gg,{\ln}_{\it 0}}^{(1)} + 8 C_A= - 8 C_A\ln\left(\frac{\muf^2}{Q^2}\right) \,,
\nn
\Delta^{(1)}_{{\cal D}_1} =& -  \Delta_{gg,{\ln}_{\it 1}}^{(1)} = ~16 C_A \,.

\end{align}
Here $\widehat{\sigma}_{(\mathrm{virt,red})}$ is the ``virtual reducible'' contribution arising from Feynman diagrams 
involving two quark triangles which in the infinite top-mass limit takes the form \cite{Altenkamp:2012sx}, 
\begin{align}
\widehat{\sigma}_{(\mathrm{virt,red})} =
\int d\widehat{t} & \, 
\as(\mur^2) \frac{4}{3} \left( \frac{\sqrt{\pi} \as(\mur^2)\alpha}{16 \sw^2 \cw^2}\right)^2
\frac{1}{\widehat{s}\MZ^4}  

\nn
\times \Bigg\{
&\MH^2   \left( 
                        -1  
                        + \frac{\MZ^2}{\widehat{t} - \MZ^2} 
                        + \ln\left( \frac{-\widehat{t}}{\MZ^2}\right) \frac{\MZ^2}{\widehat{t}-\MZ^2}
                        - \ln\left( \frac{-\widehat{t}}{\MZ^2}\right) \frac{\MZ^4}{\left(\widehat{t}-\MZ^2\right)^2}
                    \right)
                    \nn
+ &(\widehat{s}-\MZ^2) \left(
                                    -1 
                                    - \frac{\MZ^2}{\widehat{t} - \MZ^2} 
                                    +  \ln\left( \frac{-\widehat{t}}{\MZ^2}\right) \frac{\widehat{t}\MZ^2}{\left(\widehat{t}-
				    \MZ^2\right)^2}
                                    \right)
+(\widehat{t} \leftrightarrow \widehat{u})
\Bigg\} \,,
\end{align}
where $\widehat{u}$ and $\widehat{t}$ are the partonic Mandelstam variables.

The Born coefficient in the exact theory, on the other hand, contains Feynman diagrams involving triangle and box diagrams 
with heavy top-quark dependence. It can be expressed in terms of helicity amplitudes for triangle and box diagrams as,
\begin{align}\label{eq:LO-EXACT}
\widehat{\sigma}^{(0)}_{gg}(Q^2) =&
\left( \frac{ \sqrt{\pi} \as(\mur^2) \alpha}{16 s_w^2} \right)^2 \frac{1}{8\widehat{s}^2}
\int \df \widehat{t} \sum_{\lambda_g,\lambda^{'}_g,\lambda_Z} \left|
{\mathcal M}_{\lambda_g\lambda^{'}_g\lambda_Z}^\triangle
+{\mathcal M}_{\lambda_g\lambda^{'}_g\lambda_Z}^\Box\right|^2 \,.

\end{align}
The exact expression for these helicity amplitudes for two gluons and a $Z$ boson can be found in \cite{Kniehl:2011aa}.
These large distributions appearing in $\Delta^{\rm SV}$ can be resummed to all 
orders in the threshold limit $z \to 1$. Typically, resummation is performed in the Mellin $N$-space,
where plus distributions become simple logarithms in the Mellin variable ($N$). The threshold limit $z \to 1$
translates into the $N \to \infty$ limit. Recently a formalism has been proposed 
\cite{AH:2020iki, AH:2021kvg, AH:2021vdc, vanBeekveld:2021hhv, vanBeekveld:2021mxn} to also resum the NSV logarithms 
arising out of the diagonal channel. 
The formalism is built upon a factorization framework that includes a process-dependent hard function, mass factorization kernels, and a soft function, as detailed in \cite{AH:2020iki}. It closely parallels the physical kernel approach of \cite{Moch:2009hr}, wherein a single logarithmic enhancement in the large-$z$ limit was identified, persisting at all orders in $(1-z)$ near the threshold up to N3LO. This observation motivated the conjecture that such behavior continues to all orders in the strong coupling. Exploiting this conjecture, the authors of \cite{AH:2020iki} established an all-order structure for the soft function. While the structure can be explicitly verified at NLO, the higher-order coefficients are determined using the known fixed-order results for the Higgs \cite{Anastasiou:2015vya,Mistlberger:2018etf} and DY \cite{Duhr:2020seh} production processes.
In this formalism, the partonic SV and NSV coefficients in the Mellin space can be organized as follows,
\begin{align}\label{eq:resum-partonic}
	\frac{1}{\widehat{\sigma}^{(0)}_{ab}(Q^2)}Q^2\frac{\df \widehat{\sigma}_{N,ab}^{\overline{\rm N{\it n}LL}}}{\df Q^2}
	= \int_0^1 \df z ~ z^{N-1} \left(  \Delta^{\rm SV}_{ab}(z) + \Delta^{\rm NSV}_{ab}(z)  \right)
	\equiv g_{0}(Q^2) \exp \left( G^{\rm SV}_N + G^{\rm NSV}_N \right)\,.
    
\end{align}
The factor $g_{0}$ is independent of the Mellin variable and contains process-dependent information.
The leading threshold-enhanced large logarithms and the next-to-soft logarithms in Mellin space are resummed through 
the exponents $G^{\rm SV}_{N}$ and $G^{\rm NSV}_{N}$ respectively.
The resummed accuracy is determined through the 
successive terms from the exponent $G_N$, which takes the form,
\begin{align}\label{eq:GN}
G^{\rm SV}_N &= 
	\ln (\Nbar) ~g_1(\omega)
        + \sum_{n=1}^{\infty} \as^{n-1}(\mur^2)~ g_{n+1}(\omega) \,,
 \nn
G^{\rm NSV}_N &= 
        \frac{1}{N}
        \sum_{n=0}^{\infty} \as^{n}(\mur^2)~ \left( \gbar_{n+1}(\omega) + \sum_{k=0}^{n} h_{nk}(\omega) \ln^k \Nbar \right)\,,
        
\end{align}
where $\Nbar = N \exp (\gamma_E)$ with $\gamma_E$ being the Euler–Mascheroni constant and $\omega = 2 \beta_0 \as(\mur^2) \ln \Nbar$.
The first term in the expansion of $G^{\rm SV}_N$ corresponds to the leading logarithmic (LL) accuracy, whereas 
the inclusion of successive terms defines higher accuracies.
These coefficients ($g_n$) are universal and only depend on the partonic flavors being 
either quark or gluon. In order to obtain $\overline{\rm LL}$ accuracy in NSV resummation, one has to consider NSV 
exponents $\overline{g}_1$ and $h_{00}$ in addition to the LL terms. 
Similarly, for higher accuracies, one has to consider the next terms in the expansion of $G_N$. 
Note that starting from NLL ($\overline{\rm NLL}$), one has to also consider the $N$-independent $g_0$ coefficients 
whose perturbative expansion takes the form,
\begin{align}
	g_0(Q^2) = 1+\sum_{n=1}^{\infty} \as^n(\mur^2) ~g_{_{0n}}(Q^2) \,.
    
\end{align}
The explicit form of the resummed exponent for SV and NSV resummation can be 
found, \textit{e.g.}, in \cite{Catani:2003zt,Moch:2005ba}.
The SV part of the resummed result obeys the maximally non-Abelian property up to third order, and a generalized form of this property holds beyond that order. In \cite{AH:2020iki}, it was further argued that, up to NNLO accuracy, the NSV component of the soft function is universal for both DY- and Higgs-type processes. We exploit this universality to extract the $\overline{\text{NLL}}$ ingredients from the known results for Higgs production \cite{AH:2021vdc}. This implies that the soft (and next-to-soft) gluon contributions at the threshold for the gluon fusion–induced $ZH$ production at NLL ($\overline{\text{NLL}}$) accuracy can be obtained  \cite{DelDuca:2017twk} from those of gluon fusion Higgs production, once the respective Born-level and process-dependent virtual contributions are appropriately removed and the Higgs mass is treated as an arbitrary parameter.
Up to NLL($\overline{\rm NLL}$) one requires the following coefficients:
\begin{align}
    g_1(\omega) &= 
    \frac{1}{\beta_0} \bigg\{ 
                             C_A \left( 8 \left( 1+  \frac{\wb}{\omega}\Lw \right) \right)
                    \bigg\}
    \,,
    \nn
    g_2(\omega) &= \frac{1}{\beta_0^3} \bigg\{ 
            C_A \beta_1 \left( 4 \omega + 4 \Lw + 2 \Lw^2\right)
            + C_A n_f \beta_0 \left(\frac{40}{9} \left( \omega + \Lw\right) \right)
            \nn
            &\hspace{2.5em}+ C_A^2 \beta_0 \left( 
                                    \left(-\frac{268}{9} + 8 \zeta_2 \right) \left(\omega + \Lw\right) 
                            \right) 
            + C_A \beta_0^2 \left( 4 \Lw \ln  \left(\frac{Q^2}{\mur^2}\right) + 4 \omega \ln\left( \frac{\muf^2}{\mur^2}\right) 
 \right)
    \bigg\},
    \nn
    \overbar{g}_1(\omega) &= \frac{1}{\beta_0} \bigg\{ 
                                                C_A \left( 4 \Lw \right) 
                                                \bigg\} \,,
    \nn
    \overbar{g}_2(\omega) &= \frac{1}{\wb \beta_0^2} \bigg\{ 
            C_A C_F n_f \left( -8( \omega + \Lw)\right)
            + C_A^2 n_f \left( - \frac{40}{3} (\omega + \Lw)\right)
            + C_A^3 \left( \frac{136}{3} (\omega + \Lw)\right)
            \nn
            &+ C_A n_f \beta_0 \left( \frac{40}{9} \omega\right)
            + C_A^2 \beta_0 \left( - \frac{268}{9} + 8 \zeta_2\right) \omega
            + C_A \beta_0^2 \left( -8  + 4\ln \left( \frac{Q^2}{\mur^2}\right) - 4 \wb \ln\left( \frac{\muf^2}{\mur^2}\right) 
	    \right)
    \bigg\} \,,
    \nn
    h_{00}(\omega) &=  \frac{1}{\beta_0} \bigg\{
                                        C_A \left( -8 \Lw\right)
                                        \bigg\} \,,
    \nn
    h_{10}(\omega) &= \frac{1}{\wb\beta_0^2} \bigg\{ 
                                         C_A \beta_1 \left( -8 (\omega + \Lw)\right)
                                        + C_A n_f \beta_0 \left( - \frac{80}{9} \omega\right)
                                        + C_A^2 \beta_0 \left( \frac{536}{9} - 16\zeta_2 \right)\omega
\nn
&\hspace{4em}                            + C_A \beta_0^2 \left( 8 - 8\ln \left( \frac{Q^2}{\mur^2}\right)  + 
	8\wb \ln \left( \frac{\muf^2}{\mur^2}\right)  \right)  
    \bigg\} \,,
    \nn
    h_{11}(\omega) &= \frac{1}{\wb^2\beta_0} \bigg\{
                                                C_A^2 \Big( 32 \omega \wb^2 - 4 \omega\Big)                        
                                            \bigg\} \,,
                                             
\end{align}
where $\wb = 1-\omega$ and $\Lw = \ln(\wb)$.
Note that up to the required order $\NLL$, the SV and NSV exponents are the same as the $ggH$ case.
Up to $\text{NLL}~(\NLL)$ accuracy, one also needs the process-dependent coefficient $g_{_{01}}$ which for the gluon subprocess is found  to be,
\begin{align}\label{eq:g01}
{g}_{_{01}}(Q^2) &=
\left(\frac{56}{27} + 16 \zeta_2\right)C_A +\frac{64}{9}T_F\;n_f
-2 \beta_0        \ln\left(\frac{\muf^2}{\mur^2}\right)
+ \left(\frac{46}{9} C_A - 2 \beta_0 \right) \ln\left(\frac{\muf^2}{Q^2}\right)
\nn
&\hspace{1em}+\frac{\widehat\sigma_\text{(virt,red)}}{\as(\mur^2)\siglogg(Q^2)}\,.

\end{align}
Note that both $g_0$ and $G_{N}$ are scheme dependent which is related to the 
ambiguity in exponentiation of certain constant terms (e.g.\ $\gamma_E$) coming from Mellin transformation along with 
the large-$N$ terms (see e.g.\ \cite{AH:2020cok} for a detailed discussion).
In the context of the LHC, it has been observed that the so-called $\overbar{N}$-scheme provides a faster perturbative 
convergence for the resummed series. In this scheme, the constant $g_0$ is independent of 
$\gamma_E$.

The resummed result in \eq{eq:resum-partonic} has to be finally matched with the available fixed-order results to 
incorporate the hard regular contribution and, at the same time, avoid double counting of SV(NSV) logarithms.
The matching with the fixed order is usually performed using the \textit{minimal prescription} \cite{Catani:1996yz} and 
for $\overline{\rm N{\it n}LL}$ resummation it reads,
\begin{align}\label{eq:MATCHING}
	Q^2\frac{\df \sigma^{\rm N{\it n}LO+\overline{N{\it n}LL}}_{ab}}{\df Q^2}
	=&
	Q^2 \frac{\df {\sigma}^{\rm N{\it n}LO}_{ab} }{\df Q^2}
	+
        \sum_{ab \in \{gg, q\bar{q}\} }
        \widehat{\sigma}^{(0)}_{ab}(Q^2)
	\int_{c-i\infty}^{c+i\infty}
	\frac{\df N}{2\pi i}
	\tau^{-N}
	f_{a,N}(\muf)
	f_{b,N}(\muf)
	\nn
	&\times 
	\Bigg( 
		Q^2\frac{\df \widehat{\sigma}_{N,ab}^{\overline{\rm N{\it n}LL}}}{\df Q^2} 
		- 	
		Q^2\frac{\df \widehat{\sigma}_{N,ab}^{\overline{\rm N{\it n}LL}}}{\df Q^2} \Bigg|_{\rm tr}
	\Bigg) \,.
    
\end{align}
Note that a similar matching procedure is done for the SV resummation.
The $f_{a,N}$ are the Mellin-transformed 
PDFs similar to the partonic coefficient in 
\eq{eq:resum-partonic} and can be evolved e.g.\ 
using the publicly available code QCD-PEGASUS \cite{Vogt:2004ns}. 
However, for practical purposes, it can also be approximated 
by directly using the $z$-space PDF following 
\cite{Catani:2003zt,Catani:1989ne}.
The subscript `$\text{tr}$' in the last term in the brackets in \eq{eq:MATCHING} denotes that the resummed 
partonic coefficient in \eq{eq:resum-partonic} 
has been truncated 
to the fixed order to avoid double counting the  terms already present in the fixed order
through	$\sigma_{ab}^{\rm N{\it n}LO}$. Essentially, for SV resummation, this will contain all the fixed-order 
singular logarithms and for NSV resummation, it will contain additional NSV terms from the diagonal channel.
In the next section, we study the impact of SV and NSV resummation on the gluon subprocess at the LHC.
\section{Numerical Results}\label{sec:numerics}
In this section, we present numerical results for $ZH$-associated production at the LHC. The default center-of-mass energy 
of the incoming protons is set to $13.6$ TeV. Unless specified otherwise, our numerical analysis employs the 
{\tt PDF4LHC21\_40} \cite{PDF4LHCWorkingGroup:2022cjn} parton distribution functions (PDFs) throughout, as provided by 
{\tt LHAPDF} \cite{Buckley:2014ana}. In all these cases, the central set is the standard choice. The strong coupling is 
provided through the {\tt LHAPDF} routine. The fine structure constant is taken as $\alpha \simeq 1/127.93$. The masses 
of the weak gauge bosons are set to be $M_Z = 91.1880$ GeV and $M_W = 80.3692$ GeV \cite{ParticleDataGroup:2024cfk} with 
the corresponding total decay widths of the $Z$ boson, $\Gamma_Z = 2.4955$ GeV. The Weinberg angle is then given by 
$\text{sin}^2\theta_\text{w} = (1 - m_W^2/m_Z^2)$.
This corresponds to the weak coupling $G_F \simeq 1.2043993808\times 10^{-5} \text{ GeV}^{-2}$.
The mass of the Higgs boson is set to $M_H = 125.2$ GeV. 

In the gluon fusion channel, only the top-quark contribution is considered at the lowest order, with the top-quark pole 
mass set to $m_t = 172.57$ GeV.
We also chose the pole mass of the bottom quark to be $m_b = 4.78$ GeV. The unphysical renormalization 
($\mur$) and factorization scales ($\muf$) are set to the invariant mass ($Q$) of the $ZH$ pair. The scale uncertainties 
are estimated by simultaneously varying these unphysical scales in the range $[Q/2, 2Q]$ keeping the constraint 
$|\,\text{ln}(\mu_R/\mu_F)\,| < \text{ln}\,4$ (known as the seven-point scale uncertainty) and taking the maximum absolute 
deviation of the cross section from that obtained with the central scale choice. To estimate the impact of the higher-order corrections from FO and resummation, we define the following ratios of the cross sections:
\begin{align}
	{ K}_{nm}
	= 
	\frac{\sigma^{\text{N}{\it n}\text{LO}}}{\sigma_c^{\text{N}{\it m}\text{LO}}} 
	,
	{R}_{nm} 
	= 
	\frac{\sigma^{\text{N}{\it n}\text{LO} + \text{N}{\it n}\text{LL}}}{\sigma_c^{\text{N}{\it m}\text{LO}}} \text{ and }
	{ \overbar{R}}_{nm} 
	= 
	\frac{\sigma^{\text{N}{\it n}\text{LO} + \overline{\text{N}{\it n}\text{LL}}}}{\sigma_c^{\text{N}{\it m}\text{LO}}} \,.
	\label{eq:ratio}
\end{align}
The subscript `$c$' in the above expressions indicates that the corresponding quantity is evaluated at the central scale 
choice. As stated earlier, the lowest-order process for $ZH$ production through the gluon fusion channel contributes at 
$\mathcal{O}(\alpha_s^2)$ level, which formally should be considered at N2LO in the perturbation theory. Consequently, 
we have used the N2LO PDF for the computation of LO and higher-order corrections to the gluon fusion process. 

\subsection{Invariant mass distribution}
\begin{figure}[H]
\centerline{
\includegraphics[width=7.5cm, height=7.5cm]{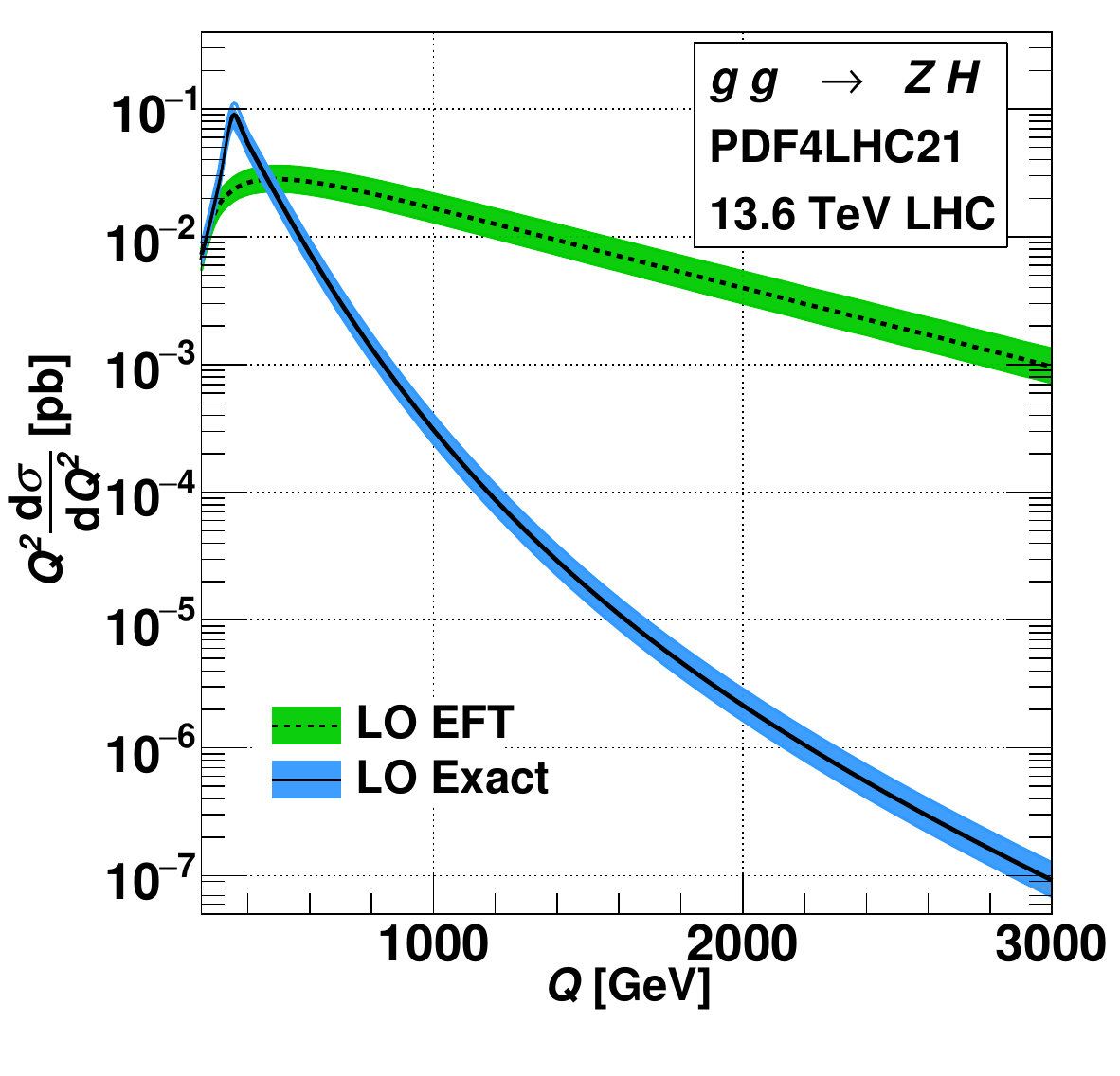}
}
\vspace{-2mm}
\caption{\small{Invariant mass distribution of $ZH$ for the gluon fusion channel at LO for EFT and Exact theories 
	at $13.6$ TeV LHC are presented. The bands correspond to the theoretical uncertainty using seven-point scale variation. }}
\label{fig:fo_kfac_eft_gg}
\end{figure}
We first look into the invariant mass distribution of the $ZH$ pair.
As mentioned in \sect{sec:introduction}, the computation of higher-order corrections is nontrivial
due to the involvement of multiple scales and one can work with EFT to reduce this complexity. For inclusive production of the
$ZH$ pair, the total cross section between exact theory and EFT differs 
by around $31\%$ at LO. However, the difference is
more pronounced in the invariant mass distribution
as can be seen from \fig{fig:fo_kfac_eft_gg}. The band in this plot corresponds to the seven-point scale uncertainty, as 
described above. Although the differential cross section is comparable in the low invariant mass region, 
the shape differs significantly in the higher invariant mass region. 
It is worth noting that in the exact theory at LO, the contribution from the triangular graphs 
is larger than that of the box diagrams in the low-$Q$ ($< 700$ GeV) region, however these two contributions are comparable to each other in the higher-$Q$ region.
However, the interference between the triangular and box diagrams
is negative and is non-negligible even in the high-$Q$ region.
Therefore, it is essential to include the finite 
quark mass effects to produce the correct behavior for the invariant mass distribution. 
On the other hand, the total cross section with finite top-quark mass effect at NLO differs only by $5\%$ \cite{Chen:2022rua} from the same in EFT. 
In fact, the exact LO captures the shape of the distribution in bulk and the effect of top quark mass on the NLO correction 
will have less impact on the shape of the invariant mass distribution
\footnote{This holds true up to a certain invariant mass region around $1$ TeV for the 13.6 TeV LHC. 
At NLO in the gluon fusion channel, with top-quark dependence,
there are also diagrams where a $Z$ boson is radiated from the external light-quark line. Such diagrams contribute at $\mathcal{O}(\alpha_s^3 y_t^2)$
($y_t$ being the top-quark Yukawa coupling), which is the same order as the gluon fusion subprocess at exact NLO.
Their contributions increase from about 1 TeV and could be 
as large as 10 times \cite{Degrassi:2022mro} the LO results at higher invariant mass region.
However, within the fiducial volume, their impact on the invariant mass distribution is less pronounced.}.
Therefore, working with a Born-improved theory where the exact LO is 
rescaled by the NLO K-factor calculated in the EFT, will reduce the complexity while retaining essential features of QCD corrections, in particular the effects of soft radiation, which we are interested in. 
It is worth noting that although the soft gluon effects themselves are unaffected by the top-quark mass, the latter can still influence the overall distribution. This dependence enters through the process-dependent virtual corrections (encoded in the hard function) as well as through the regular terms beyond the threshold SV/NSV contributions via the fixed-order matching procedure. To evaluate the accuracy of our approximation, we compared the Born-improved NLO cross section with the fully top mass–dependent result of Ref.~\cite{Chen:2020gae}, using identical input settings. We found reasonably good agreement in both the low- and high-$Q$ regions, while deviations of up to about $20\%$ appear in the intermediate-$Q$ range around $1$ TeV.\footnote{We thank Stephen Jones for providing benchmark data points for this comparison.}
In the rest of the article, we use this exact Born-improved EFT NLO cross section for all purposes.

The fixed-order results have 
been computed with the publicly available code \texttt{vh@nnlo} \cite{Harlander:2018yio,Brein:2012ne,Altenkamp:2012sx} 
and, for resummation, we have developed an in-house code to handle the Mellin inversion described at the end of 
 \sect{sec:theory}. 
 We first validated our code by reproducing the known results for the total resummed cross section of $ZH$ production up to NLO+NLL accuracy in the $N$-space formalism \cite{Harlander:2014wda}. Specifically, we adopted the same parameter choices as in \cite{Harlander:2014wda}, including the renormalization and factorization scales, parton distribution functions, prescription for the inverse Mellin transform, and the Landau-pole regularization scheme. We found excellent agreement with the published results, within the expected numerical integration uncertainties.
We further reproduced the known result for the
inclusive resummed Higgs cross section up to NLO+NLL both in $N$- and $\Nbar$-schemes. As discussed in \sect{sec:theory}, 
the $\Nbar$-scheme offers a faster perturbative convergence with a better control on the scale uncertainty, and we chose 
to use this scheme for all of our studies in this article.

\begin{figure}[ht!]
\centerline{
\includegraphics[width=7.5cm, height=7.5cm]{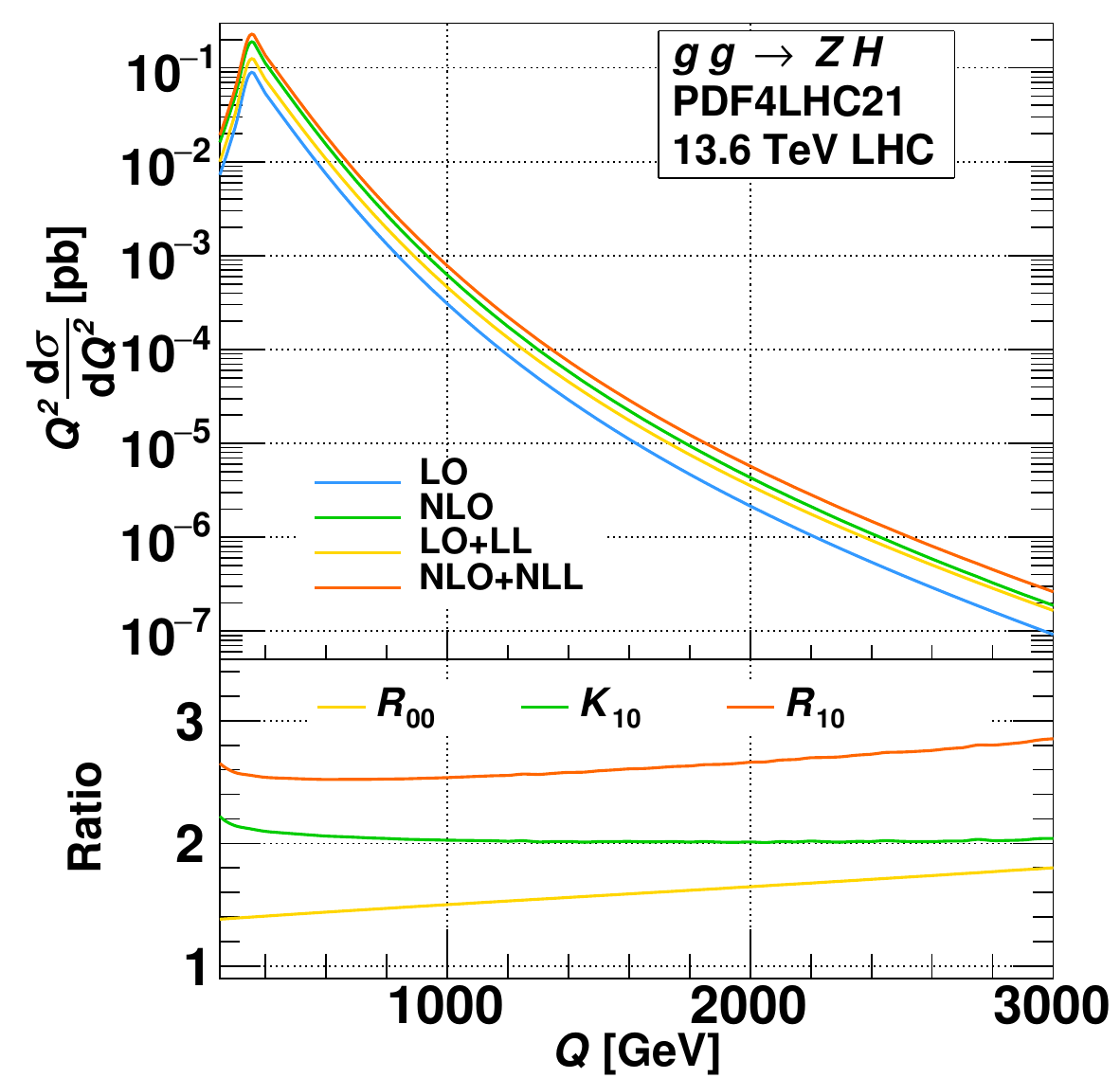}
\includegraphics[width=7.5cm, height=7.5cm]{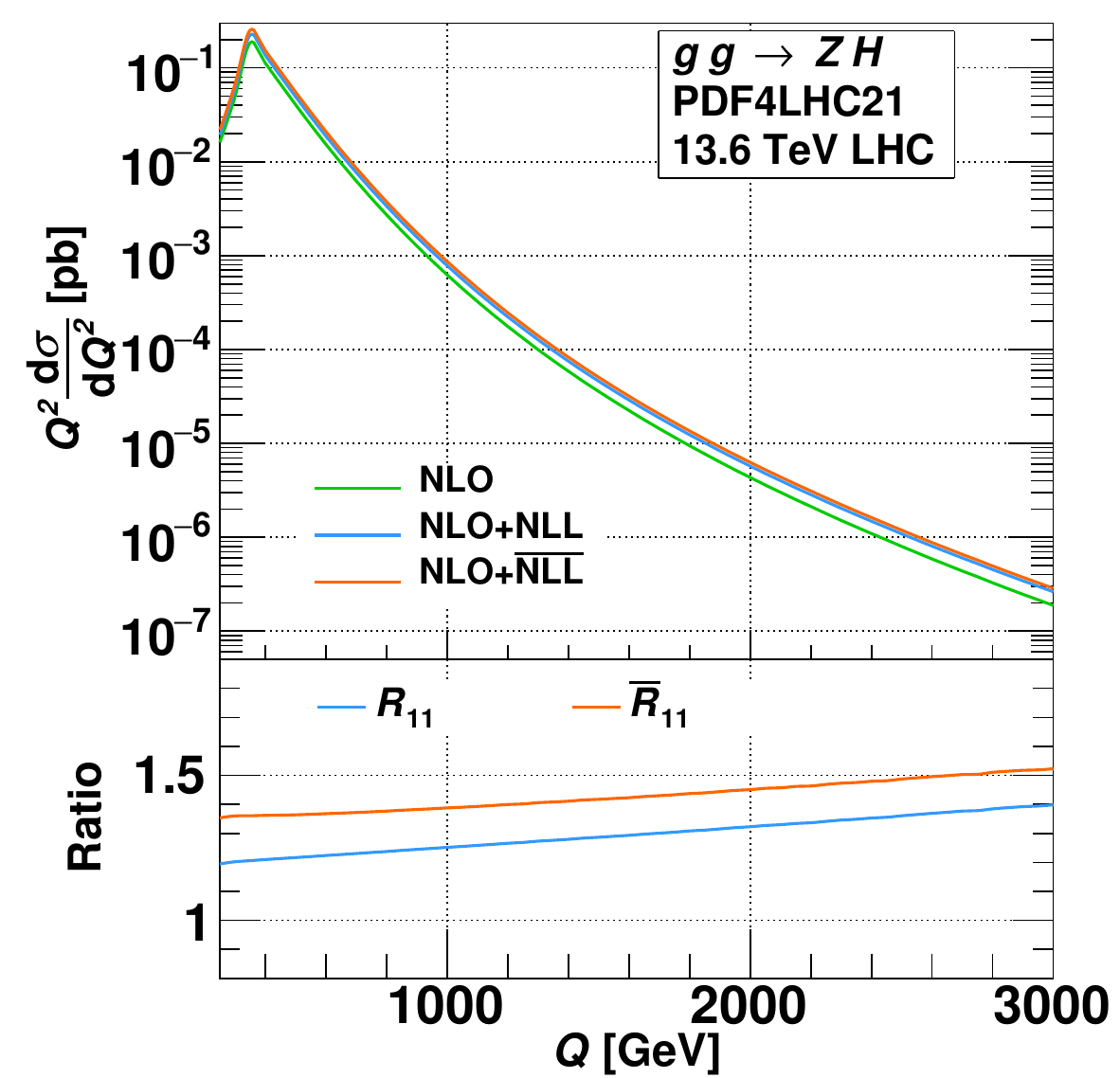}
}
\vspace{-2mm}
\caption{\small{Comparison between the Born-improved fixed order and resummed results with corresponding ratios as 
	defined in \eq{eq:ratio} are shown here. In the left panel; a comparison between fixed order and SV resummation 
	are shown, and in the right panel; a comparison among fixed order, SV resummation and NSV resummation are presented.}}
\label{fig:SV_NSV_gg}
\end{figure}
In the left panel of \fig{fig:SV_NSV_gg}, we compare the SV resummation with the fixed-order result up to NLO(NLL) 
level, and we observe the expected behavior of better perturbative convergence for the SV resummed series. This is 
further evidenced from the lower inset where the ratios are displayed. From the ratio $K_{10}$, it is evident that 
the NLO correction can reach to around $100\%$ compared to LO across most of the kinematic region considered.
The LO+LL correction captures a significant portion of the higher-order effects and provides an additional contribution 
of $80\%$ of the LO, particularly in the high invariant mass region ($Q=3000$ GeV). This clearly demonstrates that the 
soft gluon effect is dominant for this process. The NLO+NLL correction also shows a significant enhancement 
(through the $R_{10}$ factor), reaching around $2.8$ times the LO, particularly in the high invariant mass region. 
In the right panel of \fig{fig:SV_NSV_gg}, we further compare the NLO(NLL) results 
with the new $\NLL$ results. Through the ratios $R_{11}$ and $\overbar{R}_{11}$ in the bottom panel there, we observe 
that NLO+NLL corrections account for about $20\%$  of NLO in the low invariant mass region and rise to approximately 
$40\%$ in the high invariant mass region. The NSV resummation at NLO+$\NLL$, on the other hand,
contributes an additional $12\% - 15\%$ correction over NLO+NLL across most of the invariant mass region.

\begin{figure}[ht!]
\centerline{
\includegraphics[width=7.5cm, height=7.5cm]{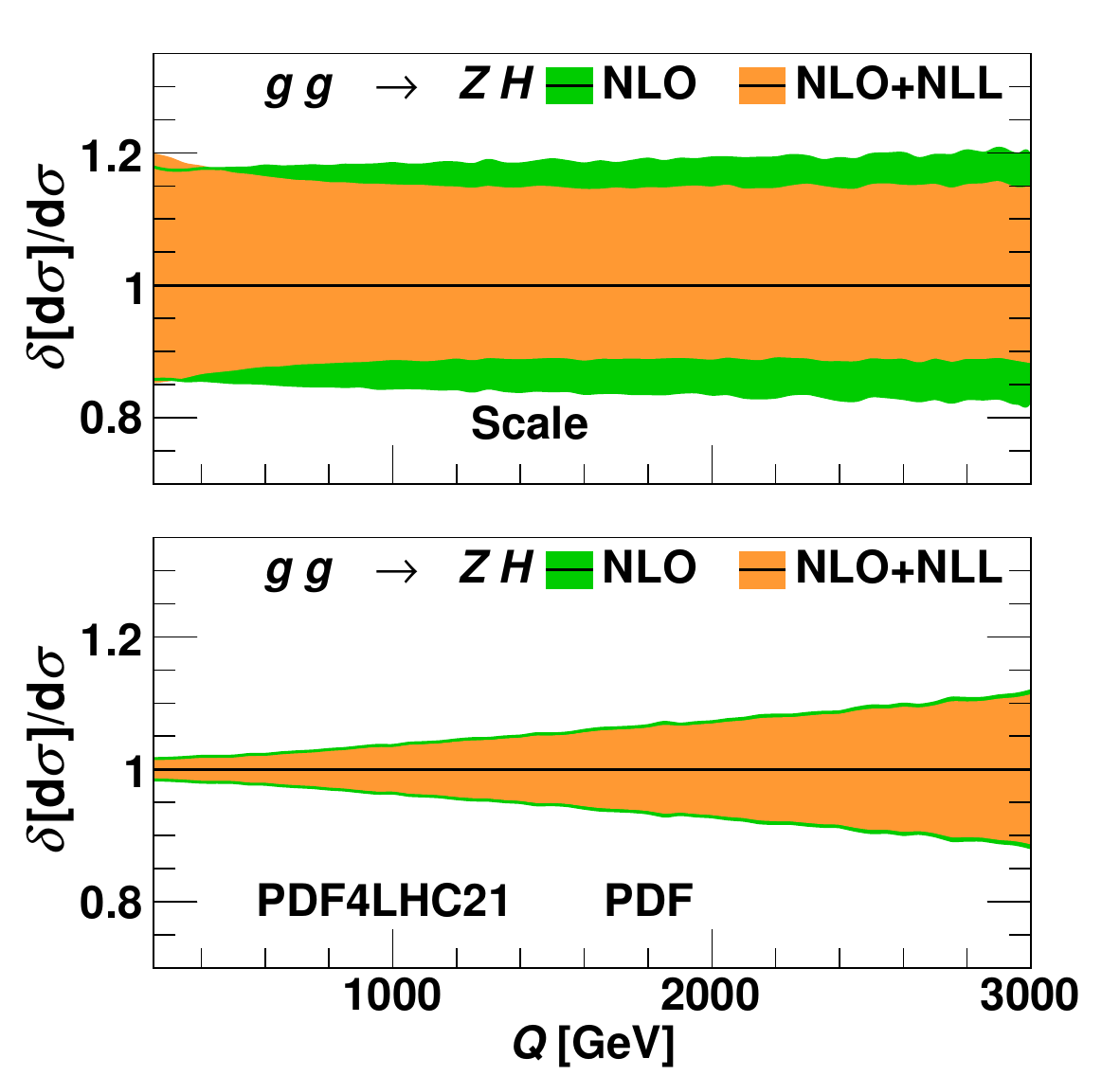}
\includegraphics[width=7.5cm, height=7.5cm]{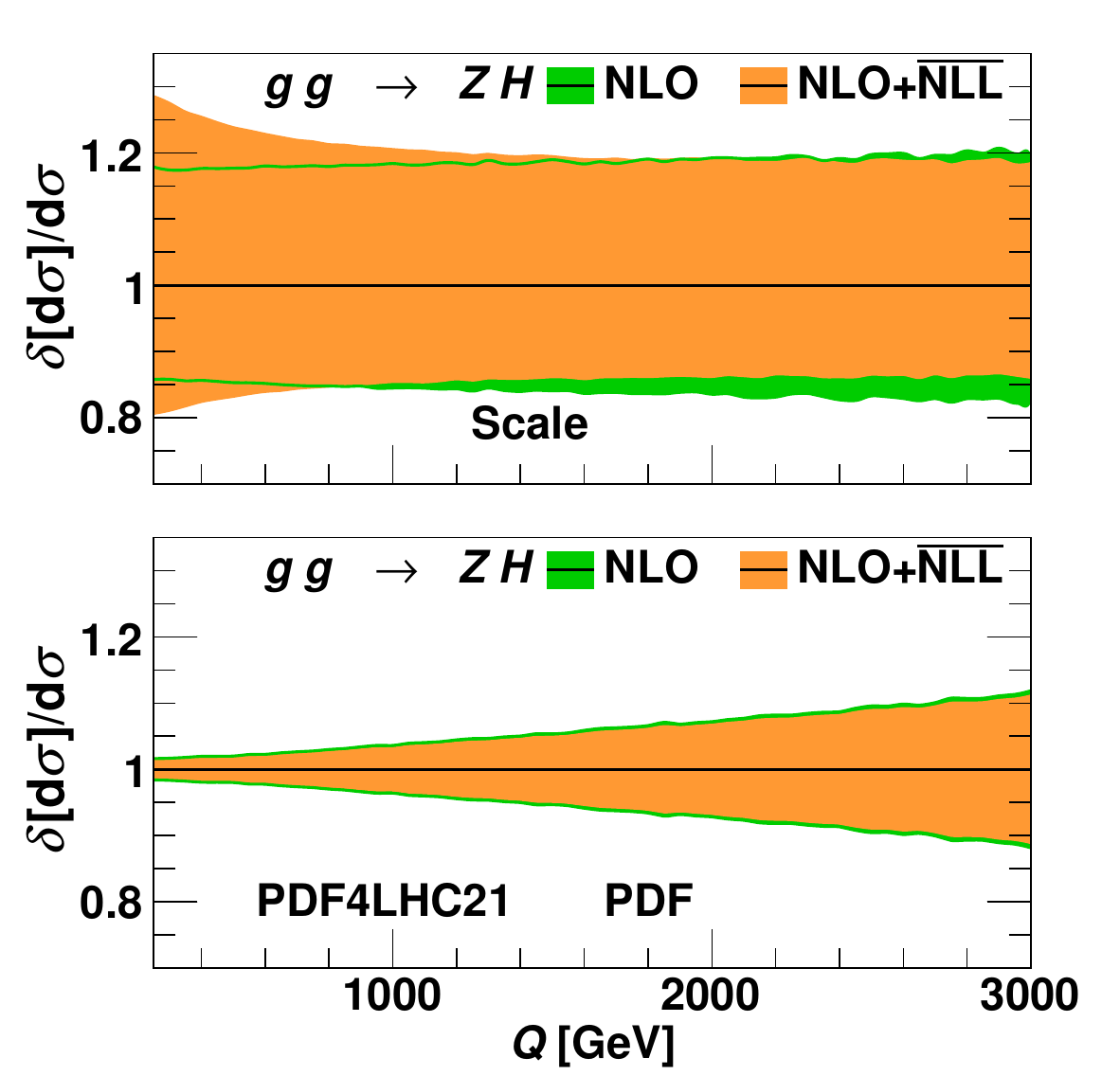}
}
\vspace{-2mm}
\caption{\small{The 7-point scale uncertainty (upper panel) and the PDF uncertainty (lower panel) of gluon fusion 
	$ZH$ production are compared for Born-improved SV (left) and NSV resummation (right) against their corresponding 
	fixed order results at $13.6$ TeV LHC.}}
\label{fig:scale_gg}
\end{figure}
In \fig{fig:scale_gg}, we have also analyzed different sources of uncertainties in these predictions.
In general, the scale uncertainties at NLO are around $20\%$, which gets reduced to around $15\%$ in NLO+NLL 
distribution, particularly in the high invariant mass region as shown in the left panel of \fig{fig:scale_gg}.
In contrast, the NSV resummation at NLO+$\NLL$ does not show similar scale reduction over NLO, particularly in 
the low invariant mass region. In the high invariant mass region, it marginally improves the scale uncertainty 
over the NLO results. In the bottom panels of \fig{fig:scale_gg}, we compare the PDF uncertainties for both SV and 
NSV resummed cases against the fixed order. These uncertainties generally increase with $Q$ as PDFs are well constrained 
in the intermediate-$x$ region relevant to the $ZH$ threshold production, whereas
at large-$x$, these are poorly constrained resulting in significant uncertainties of around $10\%$ at NLO in the high-$Q$ 
region. The SV and NSV resummed results show marginal improvements in PDF uncertainties by around $0.8\%$ over the NLO.

\begin{figure}[ht!]
\centerline{
\includegraphics[width=7.5cm, height=7.5cm]{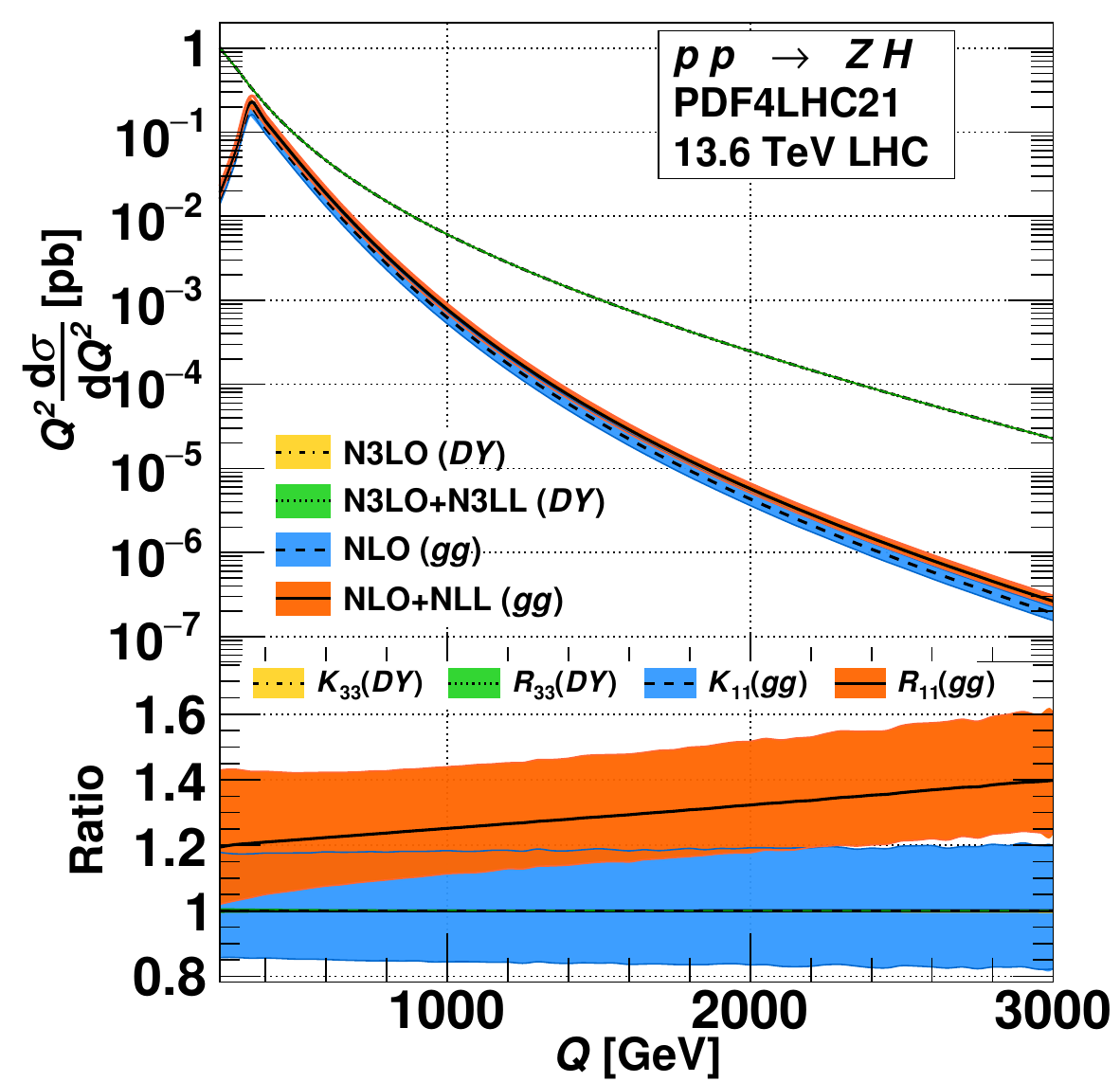}
\includegraphics[width=7.5cm, height=7.5cm]{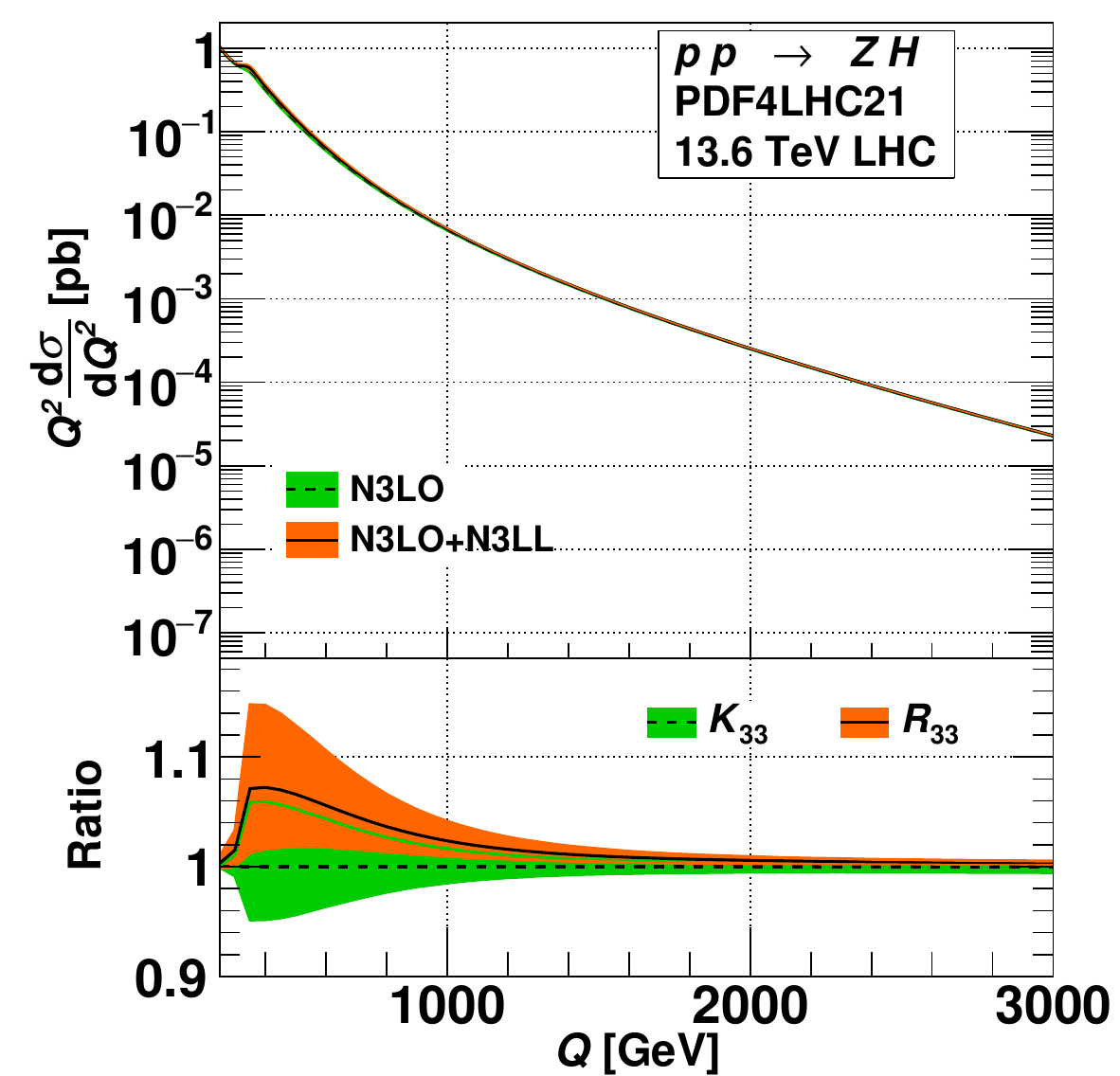}}
\vspace{-2mm}
\caption{\small{Left: The contribution from DY and gluon subprocesses are shown for fixed order and SV resummed cases 
	with the corresponding ratios at the bottom and its uncertainty from the scale variation. Right: The combined 
	contribution at the highest accuracy for the fixed order and the SV resummed order are shown along with the 
	scale uncertainties.}}
\label{fig:dis_DY_gg}
\end{figure}
\begin{figure}[ht!]
\centerline{
\includegraphics[width=7.5cm, height=7.5cm]{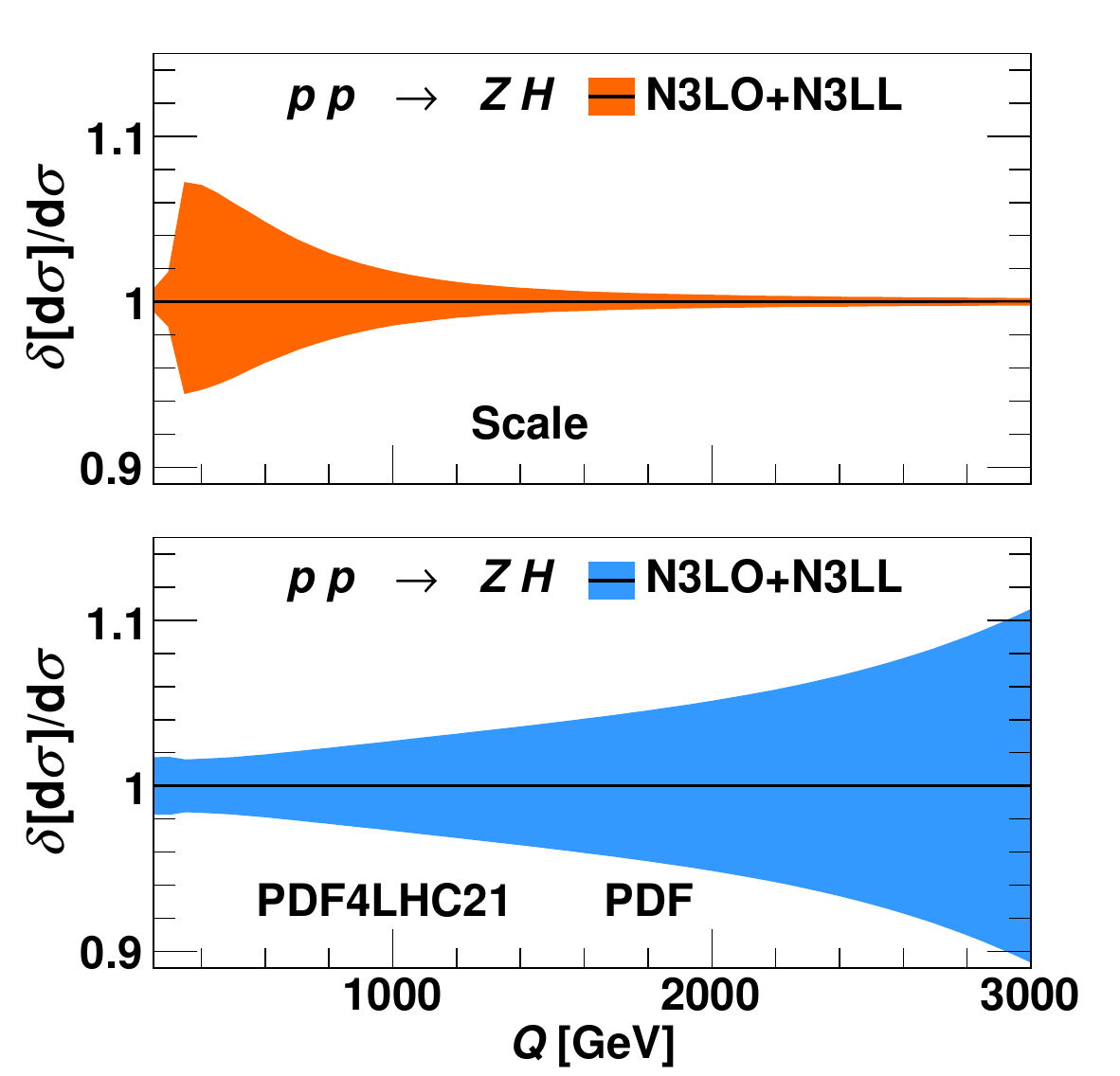}}
\vspace{-2mm}
\caption{\small{The 7-point scale uncertainty (upper panel) and the PDF uncertainty (lower panel) are shown for 
	the $pp \to ZH$ process at $13.6$ TeV LHC. }}
\label{fig:scale_ZH}
\end{figure}
\begin{align}
	Q^2\frac{\df \sigma_{pp}^{\text{N}{\rm 3}\text{LO}}}{\df Q^2} 
	&= Q^2\frac{\df \sigma_{\rm DY}^{\text{N}{\rm 3}\text{LO}}}{\df Q^2} 
	+ Q^2\frac{ \df \sigma_{gg}^{\rm NLO}}{\df Q^2} \,,
\nn
	Q^2\frac{\df \sigma_{pp}^{\text{N}{\rm 3}\text{LO+N}{\rm 3}\text{LL}}}{\df Q^2} 
	&= Q^2\frac{\df \sigma_{\rm DY}^{\text{N}{\rm 3}\text{LO+N}{\rm 3}\text{LL}}}{\df Q^2} 
	+ Q^2\frac{\df \sigma_{gg}^{\rm NLO+NLL}}{\df Q^2} \,.
	\label{eq:ZHtotaldis_resum}
\end{align}
For completeness, we now combine the gluon fusion results with other subprocesses for $ZH$ contribution to obtain full 
$pp \to ZH$ results at the $\mathcal{O}(\alpha_s^3)$ level. Particularly, in the fixed-order case, we combine the contributions 
arising from the DY-type channel with that from the gluon fusion channel at $\mathcal{O}(\alpha_s^3)$ as given in 
\eq{eq:ZHtotaldis_resum}. For thw resummation case, we chose to use the SV resummed results. While NSV resummation shows some 
enhancement over the SV resummed results (as seen from \fig{fig:SV_NSV_gg}), it still lacks contributions from off-diagonal 
channels. Moreover, one would also require the NSV accuracy for the DY-type subprocess for which the relevant ingredients 
are still missing. Hence we combine the fixed-order results with SV resummed results for both DY-type and gluon fusion 
channels. The required N3LO+N3LL SV resummed results for the DY-type channel were obtained \cite{Das:2022zie} by us a 
few years ago. In the left panel of \fig{fig:dis_DY_gg}, we show these results separately for both DY-type and gluon 
fusion channels along with the scale uncertainties. Note that the relative scale uncertainty for the DY type is not 
visible on the scale of this plot. The right panel of \fig{fig:dis_DY_gg} shows the combined results for invariant mass 
distribution at N3LO and N3LO+N3LL. The bottom panel highlights the 
enhancement from N3LO through the $K_{33}$ and $R_{33}$ factors.
In \fig{fig:scale_ZH}, we show the scale and PDF uncertainties for the combined results at N3LO+N3LL. The PDF uncertainty 
increases with $Q$, reaching around $10\%$ in the high invariant mass region. On the other hand, the scale uncertainty 
is largest at the top threshold ($Q\sim 2 m_t$), reaching up to $6\%$ and decreasing significantly in the high-$Q$ region, 
reducing to as low as $0.1\%$.

\subsection{Total cross section}
\begin{table}[ht!]
	\begin{center}{
	\setlength{\extrarowheight}{5pt}
	\scalebox{1.11}{\begin{tabular}{|c|c|c|c|c|c|}
						\hline
						Order &
						cross section (fb)  &
						Scale  &  $\delta$(PDF) & $\delta(\alpha_s)$ & $\delta$(PDF + $\alpha_s$) \\
						\hline
						\hline
						$\sigma^{\text {LO}}_{gg}$
						&  $60.25 $  &  $\pm 24.6\%$ & $ \pm 0.68 \%$
						&  $\pm 1.47 \%$ & $\pm 1.62 \%$ \\
						\hline
						$\sigma^{\text {NLO}}_{gg}$
						&  $ 124.74$ & $\pm 15.3\%$ & $\pm 0.71 \%$
						&  $\pm 1.97 \%$ & $\pm 2.09\% $ \\
						\hline
						$\sigma^{\text {LO+LL}}_{gg}$
						&  $84.74$  & $\pm 27.5 \%$ & $\pm 0.69 \%$
						&  $\pm 1.79 \%$ & $\pm 1.92\% $ \\
						\hline
						$\sigma^{\text {NLO+NLL}}_{gg}$
						&  $151.30$  & $ \pm 19.4 \%$ &  $\pm 0.71 \%$
						& $\pm 2.28 \%$ & $\pm 2.39\% $ \\
						\hline
						$\sigma^{\text {LO}+\overline{\text {LL}}}_{gg}$
						&  $96.27 $  & $\pm 29.3\%$  &  $\pm 0.69 \%$
						&   $\pm 1.95 \%$ &  $\pm 2.07\%$ \\
						\hline
						$\sigma^{\text {NLO}+\overline{\text {NLL}}}_{gg}$
						&  $170.73 $  & $ \pm 27.0\%$  & $\pm 0.71 \%$
						&  $\pm 2.51 \%$  & $ \pm 2.61 \%$ \\
						\hline
					\end{tabular}
				}}
				\caption{\small{The $ZH$ production cross sections (in fb) are presented at exact LO, 
				and Born-improved NLO, along with corresponding SV and NSV resummed results at $13.6$ 
				TeV LHC with 7-point scale, PDF and $\alpha_s$ uncertainties.}
				}
				\label{tab:tableggZH}
		\end{center}
\end{table}
Finally, we also present the total cross section for the resummed results by integrating the invariant mass distributions 
up to $\sqrt{S}$. The \tab{tab:tableggZH} provides the total cross sections for the gluon fusion channel at fixed order 
and at SV and NSV resummation. Alongside, we provide a detailed assessment of the theoretical uncertainties arising from the standard seven-point scale variation, the choice of nonperturbative PDF sets, and the variation of the strong coupling constant $\alpha_s(M_Z)$. For the latter, we follow the PDF4LHC recommendations and employ the \texttt{PDF4LHC21\_40\_pdfas} set, utilizing subsets 41 and 42, which correspond to $\alpha_s^-(M_Z) = 0.117$ and $\alpha_s^+(M_Z) = 0.119$, respectively. The central value is taken as $\alpha_s^c(M_Z) = 0.118$, consistent with the PDF set. A $1\sigma$ variation of $\pm 0.0015$ is applied to estimate the associated uncertainty, computed as
$\delta(\alpha_s) = \pm  3/4~ \big|\sigma(\alpha^+_s(M_Z))-\sigma(\alpha^-_s(M_Z))\big|/\sigma(\alpha_s^{c}(M_Z))$.
Additionally, we also present the combined uncertainties from PDF and $\alpha_s$ by adding the respective errors in 
quadrature. The largest source of uncertainty arises from the scale variation, which at the NLO+NLL level is around $19.4\%$, 
indicating the need for further improvements through higher-order computation.

\begin{table}[h!]
\begin{center}
{
{
                        	\setlength{\extrarowheight}{5pt}
	\scalebox{1.1}{\begin{tabular}{|c|c|c|c|c|c|}
								\hline
								Order &
								cross section (pb)  &
								Scale  &  $\delta$(PDF) & $\delta(\alpha_s)$ & $\delta$(PDF + $\alpha_s$) \\
								\hline
								\hline
								$\sigma^{\text {N}{\rm 3}\text {LO}}_{\rm DY}$
								&  $ 0.8416 $  &  $\pm 0.28\%$ & $ \pm 0.77 \%$
								&  $\pm 0.79 \%$ & $\pm 1.10 \%$ \\
								\hline
								$\sigma^{\text{N}{\rm 3}\text{LO+N}{\rm 3}\text{LL}}_{\rm DY}$
								&  $0.8416$ & $\pm 0.55\%$ & $\pm 0.77 \%$
								&  $\pm 0.78 \%$ & $\pm 1.10\% $ \\
								\hline
								$\sigma^{\text{N}{\rm 3}\text{LO}}_{tot}$
								&  $0.9776 $  & $\pm 1.94\%$ & $\pm 0.65 \%$
								&  $\pm 0.96 \%$ & $\pm 1.16\% $ \\
								\hline
								$\sigma^{\text{N}{\rm 3}\text{LO+N}{\rm 3}\text{LL}}_{tot}$
								&  $1.0042 $  & $ \pm 2.98\%$ &  $\pm 0.63 \%$
								& $\pm 1.03 \%$ & $\pm 1.21\% $ \\
								\hline
							\end{tabular}
						}}
						\caption{\small{The $ZH$ production cross section (in pb) are presented for DY-type and total (defined in \eq{eq:ZHtotal_resum}) at order N3LO and N3LO+N3LL for $13.6$ TeV LHC with seven-point scale, PDF and $\alpha_s$ uncertainties.}
						}
						\label{tab:tableZH}
					}
\end{center}
\end{table}

In \tab{tab:tableZH}, we present the total production cross section of $ZH$ at the LHC by combining contributions 
from different channels. Particularly, in the fixed-order case, we combine the contributions arising from the DY-type 
channel ($\sigma_{\rm DY}^{\text{N}{\rm 3}\text{LO}} $), the bottom quark-initiated $t$-channel ($\sigma_{b\bar{b}} $), 
the gluon fusion channel ($\sigma_{gg}^{\rm NLO}$), and contribution coming from Higgs radiation from top-quark loop 
in the quark annihilation channel [$\sigma_{\text{top}} (\as^2) $] as described in \sect{sec:introduction} and defined as,
\begin{align}
	\sigma_{tot}^{\text{N}{\rm 3}\text{LO}} 
	&= \sigma_{\rm DY}^{\text{N}{\rm 3}\text{LO}} 
	+ \sigma_{gg}^{\rm NLO} 
        + \sigma_{\text{top}} (\as^2) 
	+ \sigma_{b\bar{b}} \,,
    \nn
	\sigma_{tot}^{\text{N}{\rm 3}\text{LO+N}{\rm 3}\text{LL}} 
	&= \sigma_{\rm DY}^{\text{N}{\rm 3}\text{LO+N}{\rm 3}\text{LL}}  
	+ \sigma_{gg}^{\rm NLO+NLL} + \sigma_{\text{top}} (\as^2) + \sigma_{b\bar{b}} \,.
	\label{eq:ZHtotal_resum}
\end{align}
For the resummation case, we improve the fixed-order for the DY-type channel by SV resummation at N3LO+N3LL and for the 
gluon channel by resummation at NLO+NLL. For the DY-type channel, resummation has very little effect in this order and 
the major improvement is obtained through the gluon fusion channel. For the total cross section, the resummation improves 
the cross section by around $2.7\%$ compared to the fixed order, whereas the scale uncertainty increases by $1\%$.
This can be traced to the fact the total cross section receives significant contributions from the low invariant mass 
region, where the factorization scale uncertainty remains significant. However, all other sources of uncertainties remain 
small, contributing less than $1\%$.

\section{Conclusion}\label{sec:conclusion}
To summarize, we have investigated the impact of soft gluon resummation on the $ZH$ production process in the gluon 
fusion channel at the LHC. In the low invariant mass region, near the top-pair threshold $(Q \sim 2 m_t)$, the gluon 
fusion channel at the $\mathcal{O}(\alpha_s^2)$ level contributes around 20\% of the dominant DY-type channel,
highlighting its significance and the necessity of including its contribution. 

We first obtained the Born-improved NLO corrections by rescaling the exact LO results with the mass-dependent NLO $K$-factor 
obtained from the effective theory. Using the universal cusp anomalous dimensions, splitting kernels,  we have performed 
the SV and the NSV resummation and presented numerical results for the invariant mass distribution as well as the production 
cross sections to NLO+NLL($\overline{\rm NLL}$) accuracy for the current LHC energies. The NLO corrections contribute as 
large as $100\%$ of LO for the total $ZH$ production cross section in the gluon
fusion channel. We observed that the SV (NSV) resummation contributes an additional $19.4\%$ ($35.3\%$)
at the NLL 
($\overline{\rm NLL}$) level over NLO in the low-$Q$ region. In the high invariant mass region (around $Q=3000$ GeV), the
SV(NSV) resummation reduces the seven-point scale uncertainties at NLO level by a few percent $5.0\% (1.4\%)$. Besides 
the scale uncertainties, we also quantified the uncertainties due to the PDFs on resummed results, and these were 
found to be around $1.6\%$ in the low invariant mass region. 
				
Finally, for experimental analysis, we combined the contributions from different subprocesses, including the soft gluon 
resummation effects, and presented comprehensive results for the invariant mass distribution as well as the total production 
cross sections at the LHC. We note that except for 
the uncertainties due to the nonperturbative inputs from PDFs, the theoretical uncertainties in the high invariant mass 
region are well under control. However, to advance the precision for this process further, the higher-order corrections 
beyond NLO for the gluon fusion channel will be essential.			
\section*{Acknowledgements}
We thank Pulak Banerjee for the useful discussion. G. D.\ also thanks Aparna Sankar for helpful communication. The research of G. D.\ is supported by the Deutsche Forschungsgemeinschaft (DFG, German Research Foundation) under grant  
396021762 - TRR 257 (\textit{Particle Physics Phenomenology after Higgs discovery.}).
The research work of M. C. K.\ is supported by the SERB Core Research Grant (CRG) under the project CRG/2021/005270. 
The research work of K.S. is supported by the Royal Society (URF/R/231031) and the STFC (ST/X003167/1 and ST/X000745/1).
			The computation has been performed on the \textsc{OMNI} cluster at the University of Siegen,
			and on the \textsc{Param Ishan}  cluster computing facility at IIT Guwahati.
			%



			\bibliographystyle{JHEP}
			\bibliography{ggZH}
		\end{document}